\documentclass[12pt]{iopart}
\usepackage{graphicx}
\usepackage{iopams}  

\newcommand{\be}{\begin{equation}}
\newcommand{\ee}{\end{equation}}
\newcommand{\bea}{\begin{eqnarray}}
\newcommand{\eea}{\end{eqnarray}}

\newcommand{\iGAj}{{j}}

\newcommand{\cliffconj}[1] { \bar{#1} }

\begin{document}

%-------------------------------------------------------------------------
% editorial commands: to be inserted by the editorial office
%
%\firstpage{1} \volume{228} \Copyrightyear{2004} \DOI{003-0001}
%\seriesextraline{This is the Concrete Title of this Book\br H.E. R and S.T.C. W, Eds.}
%\firstpage{1}


\title[Clifford algebra spacetime]{Generalized Minkowski spacetime with geometric algebra}

%\title[Minkowski Spacetime]{Exploring the origin of Minkowski spacetime}

%----------Author 1
\author[Chappell]{James M.~Chappell}
\address{School of Electrical and Electronic Engineering, \\ University of Adelaide, SA 5005 \\ Australia}
\ead{james.chappell@adelaide.edu.au}

%----------Author 2
%%\author{John G.~Hartnett}
%\address{Institute for Photonics \& Advanced Sensing (IPAS), and the \\ School of Physical Sciences, \\ University of Adelaide, Adelaide SA 5005 \\ Australia}
%\ead{john.hartnett@adelaide.edu.au}
%Gideon1952@protonmail.com

%----------Author 3
\author{David L. Berkahn}
\address{School of Electrical and Electronic Engineering, \\ University of Adelaide, SA 5005 \\ Australia}
%\ead{davidb@physicist.net}

%----------Author 4
%\author{Azhar Iqbal}
%\address{School of Electrical and Electronic Engineering, \\ University of Adelaide, SA 5005 \\ Australia}

%----------Author 5
\author{Derek Abbott}
\address{School of Electrical and Electronic Engineering, \\ University of Adelaide, SA 5005 \\ Australia}
%\ead{derek.abbott@adelaide.edu.au}

%\subjclass{MSC 51B20, 83A05}

\vspace{2pc}
\noindent{\it Keywords}: Minkowski spacetime, Clifford geometric algebra, Electromagnetism, time, Lagrangian

\newpage

%\keywords{Minkowski, Spacetime, Clifford geometric algebra, Special relativity, Multivector}

\date{November 5, 2015}

\begin{abstract}
We begin from the generalised eight-dimensional Minkowski spacetime structure, previously developed in Clifford geometric algebra $ C\ell(\Re^3) $. We propose that this is the correct algebraic representation for physical three-dimensional space. We find that this representation incorporates spin and helicity directly into spacetime, in a Lorentz invariant manner. 
From this foundation, based on purely algebraic arguments, we derive Minkowski spacetime, the properties of electromagnetic radiation and Maxwell's equations.  These results being achieved all without physical arguments, showing that these physical laws are actually purely geometric effects.  This approach also leads to a generalization of complex mass and proper time.  Several insight about time are produced, including an arrow of time, which ultimately becomes a five-dimensional property.  We also provide a new argument explaining the non-existence of magnetic monopoles.  We suggest that the rotational freedoms, inherent in three-dimensional physical space are an important aspect of nature, not properly addressed in physics, as they are not incorporated within the spacetime background, as achieved in this paper.
\end{abstract}

%%% ----------------------------------------------------------------------
\maketitle
%%% ----------------------------------------------------------------------
%\tableofcontents

\section{\label{sec:level1}Introduction}

Descartes first conceived the revolutionary idea of a spatial grid that could be used as a reference system to described the motion of objects, and then later, with time also being included as a absolute universal reference, akin to space, by Newton.  Minkowski, following Hamilton's ideas, generalized this framework with a unified space and time continuum, with each being interdependent, and which Einstein later further generalized to being a flexible grid which can be deformed to produce the effect of gravity.  This flexible Minkowski spacetime framework is now viewed as the fundamental backdrop within which to describe the laws physics.

Now, it has  been shown previously~\cite{chappell2023}, that a  Minkowski-type spacetime structure naturally emerges within the Clifford algebra $ C\ell(\Re^3) $, of three dimensions. This generalised eight-dimensional spacetime can be expressed in three-vector notation with the multivector
\be \label{multivector3DSimple}
d X =  d t + d \boldsymbol{x} + \iGAj d \boldsymbol{n} + \iGAj d b ,
\ee
where $ t $ is the local time, $ \boldsymbol{x} = x_1 e_1 + x_2 e_2 + x_3 e_3 $ a spatial vector, $ \iGAj \boldsymbol{n} = n_1 e_2 e_3 + n_2 e_3 e_1 + n_3 e_1 e_2 $ a bivector representing spin, where $ \boldsymbol{n} = n_1 e_1 + n_2 e_2 + n_3 e_3 $, and $ \iGAj = e_1 e_2 e_3  $ the trivector describing helicity, with $ b, x_1, x_2, x_3, n_1, n_2, n_3 $  real scalars~\cite{Chappell2015,Chappell2014IEEE}.
We have used as a basis for three dimensional Clifford's algebra $ C\ell \left ( \Re^3 \right ) $, the three unit elements $ e_1, e_2, e_3 $, having a unit square $ e_1^2 = e_2^2 = e_3^2 = 1 $, and are anticommuting, with $ e_1 e_2 = - e_2 e_1 $, $ e_1 e_3 = - e_3 e_1 $ and $ e_2 e_3 = - e_3 e_2 $. 
%The four types of elements within $ C\ell \left ( \Re^3 \right ) $ naturally describe the four distinct physical quantities typically represented as scalars, polar vectors, axial vectors (or pseudovectors) and pseudoscalars, respectively.
In order to produce a suitable invariant interval, we define the operation of Clifford conjugation
\be \label{CliffordConjugation}
d \cliffconj{X} =  d t - d \boldsymbol{x} - \iGAj d \boldsymbol{n} + \iGAj d b .
\ee
Clifford conjugation is an anti-automorphism, so that for a product $ M N $ of two multivectors $ M , N \in  C\ell \left ( \Re^3 \right ) $, $ {\overline{M N}} = \cliffconj{N} \cliffconj{M} $.  Clifford conjugation can be written algebraically as $ \cliffconj{M} = \frac{1}{2} \left (  e_1 M e_1 + e_2 M e_2+ e_3 M e_3 -M \right ) $. We note that Clifford conjugation is equivalent to a time reversal operation.

\subsubsection{The invariant interval} 
The invariant interval of a multivector $ M $ using Clifford conjugation is calculated using the bilinear form
\be \label{AmplitudeSquared}
| d X |^2 = d X d \cliffconj{X}  = d t^2 - d \boldsymbol{x}^2 + d \boldsymbol{n}^2 - d b^2 + 2 \iGAj \left (d b d t - d \boldsymbol{x} \cdot d \boldsymbol{n}  \right ),
\ee
forming a complex-like number, and thus commuting with the rest of the algebra.  
We refer to this as a `complex-like' number because, the trivector $ \iGAj $ has the properties of the unit imaginary, with $ \iGAj^2  = -1 $, and all other quantities are real scalars. The square root is therefore well defined from complex number theory and so we can define the multivector {\it amplitude} as $ | M | = \sqrt{| M |^2 } $.    We have $ |M_1 M_2|^2 = |M_1|^2 |M_2|^2 $, so that we can write a norm relation
\be
| M_1 M_2 |  = \pm | M_1| |M_2| ,
\ee
provided the appropriate branch is used when finding the complex square roots.
We thus have a distance measure between two multivectors $ M_1 , M_2 $, of $ | M_1 -M_2 |$, making $ C\ell(\Re^3) $ a metric space.  

\subsubsection{Generalised Lorentz transformations}  For transformations that are continuous with the identity, using the power series expansion of the exponential function, 
 we can write a general transformation operation
\be \label{HomLorentzGroup}
M' = \rme^{ \boldsymbol{p} + \iGAj \boldsymbol{q}  } M \rme^{ \boldsymbol{r} + \iGAj \boldsymbol{s} },
\ee
which will leave the multivector amplitude invariant. 
That is \be \label{HomLorentzGroupShowInvariant}
M' \cliffconj{M'}= \rme^{ \boldsymbol{p} + \iGAj \boldsymbol{q}  } M \rme^{ \boldsymbol{r} + \iGAj \boldsymbol{s} }  \rme^{ -\boldsymbol{r} - \iGAj \boldsymbol{s} } \cliffconj{M} \rme^{ -\boldsymbol{p} - \iGAj \boldsymbol{q}  } = M \cliffconj{M}.
\ee

The four three-vectors $ \boldsymbol{p}, \boldsymbol{q}, \boldsymbol{r}, \boldsymbol{s} $ illustrate that the set of transformations is a twelve dimensional manifold, thus generalizing the conventional six dimensional Lorentz group, consisting of boosts and rotations. 

Crucially, we note that we actually had no choice in choosing the involution of Clifford conjugation, as it is the only one having the essential property of producing a commuting quantity for the invariant interval, allowing this Lorentz-type invariance.

\subsubsection{Multivector dot product}  Now, since $ M \cliffconj{M} $ is invariant, then $ (A + B)(\overline{A + B}) $ must also be invariant, where $ A,B \in C\ell(\Re^3)$. We have
\be
(A + B)(\overline{A + B}) = A \cliffconj{A} + B \cliffconj{B} + A \cliffconj{B} + B \cliffconj{A}  .
\ee
Hence, as $ A \cliffconj{A}, B \cliffconj{B} $ are known to be invariant, then we can define a multivector dot product with the final two terms
\be \label{MultivectorDotProduct}
A \cdot \cliffconj{B} = \frac{1}{2} \left ( A \cliffconj{B} + B \cliffconj{A} \right ) = B \cdot \cliffconj{A} .
\ee
The dot product thus provides a mechanism to combine two distinct multivectors in an invariant manner, as in the electromagnetic interaction Lagrangian $ A \cdot \cliffconj{J} $, for example.

%As mentioned, the result that $ |V|^2 = 1 $ for all observers, implies that the magnitude of the eight-velocity is constant. Hence, the change of $V$ must be orthogonal to $V$, which is shown by $ A \cdot \cliffconj{V} = 0 $.  We have $ |A|^2 < 0 $, so that $U$ is getting a spacelike variation $ d U = A d \tau $. The spacelike acceleration, $ |A|^2 < 0 $, implies that the change to the temporal part is less than than the change to the spatial part, and so ensuring that the magnitude of $ V $ is constant. Hence, acceleration can be described as a rotation of the velocity multivector.
%In general, the spatial parts of V and A do not have to be orthogonal to each other but this is the case for circular motion with $ \boldsymbol{a} \cdot \boldsymbol{v} = 0 $, for example. 
%The velocity multivector thus also has the special property of being dimensionless.  %It also has a magnitude of zero for lightlike particles with $ |d X |^2 = 0 $, or unity for timelike particles $ |d X |^2 \ne 0 $.

\subsubsection{Fields} We identify a special class of multivectors that are produced from a product of two multivectors $ \cliffconj{A} B $, which will transform as
\be \label{multivectorProductCliffConj}
 \cliffconj{A}' B'  = \overline{K A L} K B L  = \cliffconj{L} \cliffconj{A} \cliffconj{ K} K B L  =   \cliffconj{L }  \cliffconj{A} B L  .
\ee
Hence multivectors formed as a product $ F = \cliffconj{A} B $ form a distinct class of multivectors with a restricted transformation law 
\be
F' = \cliffconj{L } F L .
\ee
We will refer to such quantities as {\it fields}, as it turns out that these transformations are isomorphic to the conventional Lorentz transformation for the electromagnetic field.

\subsubsection{The electromagnetic potential}

We define the four-gradient 
\be \label{FourGradient}
\partial = \frac{\partial}{c \partial t} + e_1 \frac{\partial}{\partial x} + e_2 \frac{\partial}{\partial y} + e_3 \frac{\partial}{\partial z} ,
\ee
in the conventional manner.
Now, if we define
\be
F = \cliffconj{\partial} A ,
\ee
where  we have defined an  eight-potential $ A = \frac{\phi}{c} -  \boldsymbol{A} - \iGAj  \boldsymbol{C} + \iGAj \frac{\psi}{c} $,
%, where $ \iGAj  \boldsymbol{C} + \iGAj \frac{\psi}{c} $ represent magnetic monopole potentials.
then this meets our conditions for a field, as both $ \partial , A  \in C\ell(\Re^3)$. 
For completeness, we note that we can produce a scalar wave operator
\be  \label{WaveOperator}
\partial \cliffconj{\partial} = \frac{\partial^2}{c^2 \partial t^2} - \frac{\partial^2}{\partial x^2} - \frac{\partial^2}{\partial y^2} - \frac{\partial^2}{\partial z^2} .
\ee

We thus have the expression for the field
\be
F  = \cliffconj{\partial} \boldsymbol{A} = \left ( \frac{\partial}{c \partial t} - \nabla \right ) \left ( \frac{\phi}{c} -  \boldsymbol{A} - \iGAj  \boldsymbol{C} + \iGAj \frac{\psi}{c} \right ) ,
\ee
which gives the four equations
\bea
\ell & = & \frac{\partial \phi}{c^2 \partial t} + \nabla \cdot \boldsymbol{A} ,\\ \nonumber
\boldsymbol{E}  & = & - \nabla \phi - \frac{\partial \boldsymbol{A}}{ \partial t} + \iGAj \nabla \wedge \boldsymbol{C}, \\ \nonumber
 \iGAj \boldsymbol{B} & = &  \nabla \wedge \boldsymbol{A} - \iGAj \frac{\partial \boldsymbol{C}}{c \partial t} -  \frac{\iGAj}{c} \nabla \psi , \\ \nonumber
\iGAj \kappa & = & \iGAj \left (\frac{\partial \psi}{c^2 \partial t } +  \nabla \cdot \boldsymbol{C} \right ) . \nonumber
\eea
This gives the conventional field definitions in Maxwell's equations $ \boldsymbol{E}_M = - \nabla \phi - \frac{\partial \boldsymbol{A}}{ \partial t} $ and $  \iGAj \boldsymbol{B}_M = \nabla \wedge \boldsymbol{A} = \iGAj \nabla \times \boldsymbol{A} $, as well as the Lorenz gauge $ \ell =\frac{\partial \phi}{c^2 \partial t} + \nabla \cdot \boldsymbol{A} $.  In order to recover the standard electromagnetic field $ F = \boldsymbol{E}/c + \iGAj \boldsymbol{B} $, with zero scalar and pseudoscalar components, we need to therefore adopt the Lorenz gauge with $ \ell = \kappa = 0 $. The Lorenz gauge has been shown to  produce a Lorentz invariant form of electromagnetism, enforcing causality and charge conservation, which is generally assumed to be a requirement of a physical theory.  %There also appears to be an identical conservation principle for the magnetic monopole potentials $ \iGAj  \boldsymbol{C} + \iGAj \frac{\psi}{c}$, so we also take $ \kappa = \frac{\partial \psi}{c^2 \partial t } + \nabla \cdot \boldsymbol{C} = 0 $.
%Hence, the full eight-dimensional potential multivector $ A $ can be utilised to generate the standard electromagnetic field from $ F = \cliffconj{\partial} \cliffconj{A} $, provided we adopt the Lorenz gauge.
%Alternatively, if we want to create a field $ F = \boldsymbol{E}/c + \iGAj \boldsymbol{B} $ without the need to enforce the Lorenz gauge, we can take
%\be
%\boldsymbol{E}/c + \iGAj \boldsymbol{B} = \frac{1}{2} (F - \cliffconj{F}) = \frac{1}{2} \left (\cliffconj{\partial} \cliffconj{A} - A \partial \right ) ,
%\ee
%where in the second term, the gradient operator acts back to the left on $ A $.

Now, we have Maxwell's source free equations $ \partial F = 0 $
\be
\left ( \frac{\partial}{ \partial t} + \nabla \right ) \left ( \ell + \boldsymbol{E}_M +   \iGAj \nabla \wedge \boldsymbol{C} + \iGAj   \boldsymbol{B}_M - \iGAj \frac{\partial \boldsymbol{C}}{ \partial t} - \iGAj  \nabla \psi + \iGAj \kappa \right ) = 0 ,
\ee
where for clarity we use units where $ c = 1 $.
For the scalar components we have
\bea
& & \nabla \cdot \boldsymbol{E}_M + \frac{\partial \ell}{ \partial t} +  \iGAj \nabla \wedge \nabla \wedge \boldsymbol{C} \\ & = & \nabla \cdot \boldsymbol{E}_M  +  \iGAj \nabla \wedge \nabla \wedge \boldsymbol{C}= 0 ,
\eea
assuming the Lorenz gauge is zero $ \ell = 0 $. Then, considering Maxwell's equation $ \nabla \cdot \boldsymbol{E}_M = \rho $, then the electric charge can be identified with the helicity of a monopole current potential, $ \rho = -\iGAj \nabla \wedge \nabla \wedge \boldsymbol{C} $, a scalar.  Hence, electric charges can form from the potentials in the same way as electric and magnetic fields.

For the pseudoscalar terms or magnetic monopole charge, we have 
\bea
& & \nabla \cdot \boldsymbol{B}_M + \frac{\partial \kappa}{\partial t}  -  \frac{\partial \nabla \cdot \boldsymbol{C}}{ \partial t} -   \nabla^2 \psi \\
& = &  \nabla \cdot \boldsymbol{B}_M + \frac{\partial^2 \psi}{ \partial t^2} + \frac{\partial \nabla \cdot \boldsymbol{C}}{\partial t} -\frac{\partial \nabla \cdot \boldsymbol{C}}{ \partial t} -  \nabla^2 \psi = \nabla \cdot \boldsymbol{B}_M =0 ,
\eea
because we know $ \frac{\partial^2 \psi}{ \partial t^2} -  \nabla^2 \psi =0$, as we are assuming Maxwell's source-free equations.  Hence, if there are no sources there is no way to generate a charge from the potentials, in distinction to the electric case, which can form from the helicity of the potential.  Hence, for magnetism we have $\nabla \cdot \boldsymbol{B}_M =0 $, as monopoles cannot form from the potentials.

\section{The action}

As first proven by Noether~\cite{Noether1918}, wherever there is a symmetry of nature there is an associated conservation law. The conservation laws in turn imply the apparent presence of forces. The invariant distance provides a suitable action integral
\be \label{ActionMetricDistance}
S =  \int | d X | ,
\ee
where the distance $  | d X | $ is given by the amplitude of the spacetime multivector, given by Eq.~(\ref{AmplitudeSquared}). That is, we are following the standard procedure of extremizing the proper time in order to find the geodesics. For a null distance (representing electromagnetic radiation), we have the spacetime distance
\be
|dX|^2 = \left ( \dot{t}^2 - \dot{\boldsymbol{x}}^2 + \dot{\boldsymbol{n}}^2 - \dot{b}^2  \right ) d s^2 = 0,
\ee
where we define $ \dot{t} = \frac{d t}{d s} $, $ \dot{\boldsymbol{x}} =  \frac{d \boldsymbol{x}}{d s} $, $ \dot{\boldsymbol{n}} =  \frac{d \boldsymbol{n}}{d s} $ and $ \dot{b} = \frac{d b}{d s} $ using an arbitrary time scale $ s $, as no proper time is defined for light.  We can then write the action as $ S = \int \frac{| d X | }{d s} d s $ that implies a Lagrangian
\be \label{LagrangianVelocity}
\mathcal{L} = \frac{| d X | }{d s} = |V|  = \sqrt{ \dot{t}^2 - \dot{\boldsymbol{x}}^2 + \dot{\boldsymbol{n}}^2 - \dot{b}^2 } = 0,
\ee
where we now extremize  $ S = \int \mathcal{L} d s $.

As we have no explicit coordinate dependence,  $ \frac{ \partial \mathcal{L}}{\partial \dot{t} } $, $  \frac{ \partial \mathcal{L}}{\partial \dot{\boldsymbol{x}} } $, $  \frac{ \partial \mathcal{L}}{\partial \dot{\boldsymbol{n}} } $ and $ \frac{ \partial \mathcal{L}}{\partial \dot{b} } $ are constants of the motion.
Using the Euler-Lagrange equation~\cite{goldstein2002} for $ t $ 
\be
\frac{d}{d s} \frac{ \partial \mathcal{L}}{\partial \dot{t} } = \frac{ \partial \mathcal{L}}{\partial t }  = 0
\ee
thus giving the conserved quantity
\be  \label{energyInvariantLagrangian}
\frac{ \partial \mathcal{L}}{\partial \dot{t} } =  \dot{t} = E  .
\ee
We have written the conserved quantity $ E $ as we expect it to relate to energy by Noether's theorem.  We expect the second conserved quantity will be conserved momentum $ \boldsymbol{p} = \dot{\boldsymbol{x}} $.  
The bivector component will produce the conservation of  angular momentum $ \boldsymbol{s} = \dot{\boldsymbol{n}} $ as expected and 
the fourth conserved quantity will be
\be
\frac{ \partial \mathcal{L}}{\partial  \dot{b}  } = \dot{b} = H,
\ee
that returns the helicity $ H =  \dot{b}  $.  
We thus find that the invariant interval in Eq.~(\ref{ActionMetricDistance}), encodes the four fundamental conservation laws, energy, momentum, angular momentum and helicity~\cite{lehmkuhl2014einstein}. 

Nows, the simplest extension of the inertial Lagrangian, would  be $ \mathcal{L} =  |V + U | $, where the multivector $ U $  would represent a `flow' in the background spacetime, modifying particle inertial motion $ V $, giving asuitable Lagrangian
\be \label{GravityLagrangian}
\mathcal{L} =  \frac{1}{2}| V+ U|^2 .
\ee
Note that we are permitted to use either $ \mathcal{L} =  |V + U| $ or $ \mathcal{L} =  \frac{1}{2} |V + U |^2 $, because if a Lagrangian $ \mathcal{L} $ satisfies the Euler-Lagrange equations, then in general any function  $ F(\mathcal{L}) $ of the Lagrangian also satisfies the Euler-Lagrange equations.  
The idea of an offset velocity vector $ U $ added in a Galillean manner, but with particles Lorentz boosted within this flow $ U $, thus mimics the Gullstrand–Painlevé coordinates also known as the `rain frame' of General Relativity~\cite{taylor2001exploring}.
We can include electromagnetic forces with the conventional addition $ A \cdot \cliffconj{V} $, giving
\be \label{unifiedLagrangian}
\mathcal{L} =  \frac{1}{2}| V+ U|^2  + \left (\frac{q}{m} \right ) A \cdot \cliffconj{V} .
\ee
These minimal additions to the inertial Lagrangian appear to lead to the idea of metric-type gravitational forces as well as EM type forces.  While this certainly can duplicate Newtonian gravitational and classical Maxwell electromagnetic forces, more work needs to be done to see how closely it could approximate general relativity more generally.  Hence, minimal additions to the inertial Lagrangian naturally produces the force laws for the two long range forces in nature.

\subsection{Null intervals}

\begin{figure}[htb]

\begin{center}
\includegraphics[width=4.4in]{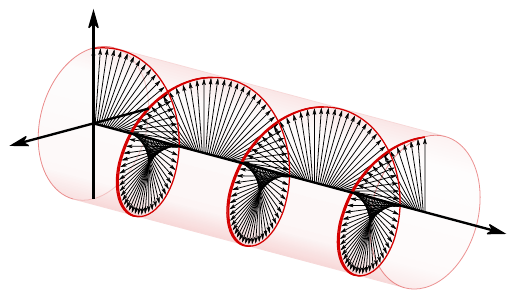}
\end{center}

\caption{Circularly polarized transversely twisting propagation (left handed). A property that is intrinsic to the algebra of $ C\ell(\Re^3) $ for null particles. \label{circularlyPolarized}}

\end{figure}

Considering the null condition $ |d X|^2 = 0 $ of the metric in Eq.~(\ref{AmplitudeSquared}), and  assuming a light speed particle with $ c^2 d t^2 - d \boldsymbol{x}^2 = 0 $, we then require firstly $ c^2 d b^2= d \boldsymbol{n}^2 $ or $ c d b= \pm || d \boldsymbol{n} || $ and secondly $ c^2 d b d t - d \boldsymbol{x} \cdot d \boldsymbol{n}=0  $. By combining these two equations we produce  the condition $ d \boldsymbol{x}\cdot d \boldsymbol{n}= \pm c || d \boldsymbol{n} || d t $. Dividing through by $ d t^2 $, we find  
\be
\boldsymbol{v}\cdot \hat{\boldsymbol{w}}= \pm c  ,
\ee
where we define the angular frequency $ \boldsymbol{w} = \frac{d \boldsymbol{n}}{d t} $ and the velocity $ \boldsymbol{v} = \frac{d \boldsymbol{x}}{d t} $.
Hence, due to the nature of the dot product, it is enforcing a spacelike condition, and we can see that it is only satisfied by a velocity $ || \boldsymbol{v} || = c $,  parallel or anti-parallel to the spin axis $ \hat{\boldsymbol{w}} $. 
Additionally, using the condition $ c^2 d b d t - d \boldsymbol{x} \cdot d \boldsymbol{n} = 0 $, we produce a constant helicity
\be \label{spacetimeHelicity}
\xi = \frac{d b}{d t} = \frac{1}{c^2} \boldsymbol{c} \cdot \boldsymbol{w} = \frac{w}{c} = \frac{2 \pi }{\lambda} = k .
\ee 
That is, based on the eight-dimensional structure of $ C\ell(\Re^3) $ alone, we find that the null interval, with a propagation at the speed of light $ c$, is required to have its spin axis parallel to its direction of propagation, with two possible helicities~(proportional to frequency), exactly as observed for circularly polarised photons, a naturally occurring form of electromagnetic radiation.  
Thus, the intrinsic oscillations over $ C\ell \left ( \Re^3 \right ) $ consists of a transverse twist as it moves through the vacuum, as shown Fig.~\ref{circularlyPolarized}.
 Helicity is a property not as well known as spin, but can be visualised as the operation of wringing a wet towel, for example.  There is no net rotation of the towel, just a twisting action at the central point.  Helicity is a phenomena that can only arise in three dimensions, whereas planar spin can exist in two dimensions, spatial displacement can exist in one dimensions, and time a scalar can exist in zero dimensions.  %Hence, in this sense, time is the more universal property of spacetime.

The wave in Fig.~\ref{circularlyPolarized}, assumed to be propagating along the $ e_3 $ or $z$-axis, can be described algebraically as
\be \label{CircularlyPolarizedEField}
F = \boldsymbol{E}_0 \rme^{\iGAj e_3 \left ( w t - k z \right )} =  E_0 e_1 \cos  \left (w t - k z \right ) + E_0 e_2 \sin \left ( w t - k z \right ) ,
\ee
where  we assume $ \boldsymbol{E}_0 = E_0 e_1 $ is a vector in the $ x-y $ plane with an orientation at $ t = 0 $ along the $ x$-axis.  The bivector $ \iGAj e_3 $, acts as a rotation operator acting on the initial vector direction $ \boldsymbol{E}_0 $, creating the twisting motion as it propagates down the $ e_3 $ axis, as shown in Fig.~\ref{circularlyPolarized}. 
However, we have initially assumed only a single vector is being twisted as it moves through space, however, more generally this could be a full multivector 
\be
F = \ell + \boldsymbol{E} + \iGAj \boldsymbol{B} + \iGAj \kappa,
\ee
which is being rotated.
We require
\be \label{generalNullRelationField}
F \cliffconj{F}  =   \ell^2 - \boldsymbol{E}^2 +  \boldsymbol{B}^2  - \kappa^2  + 2 \iGAj \left(  \ell \kappa -  \boldsymbol{E} \cdot \boldsymbol{B} \right ) = 0 .
\ee

%This can easily be generalised for an arbitrary propagation direction $ 
%F = \boldsymbol{E}_0 \rme^{\iGAj \hat{\boldsymbol{k}} \left ( w t \pm \boldsymbol{k} \cdot \boldsymbol{x}\right )}$, where $  \boldsymbol{E}_0 $ rotates in the plane orthogonal to $\hat{\boldsymbol{k}}$, and the positive and negative signs correspond to right-handed and left-handed polarization, respectively.
 %We find
%\be
%F \cliffconj{F} = -\left (\boldsymbol{E} + j c \boldsymbol{B} \right )\left (\boldsymbol{E} + j c \boldsymbol{B} \right ) = \boldsymbol{E}^2 - c^2 \boldsymbol{B}^2 - 2 c \iGAj \boldsymbol{E} \cdot \boldsymbol{B} .
%\ee
If we take the Lorenz gauge terms  $ \ell = \kappa  = 0$, then $ |\boldsymbol{E}| = |\boldsymbol{B}|$ and $ \boldsymbol{E}$ is orthogonal to $ \boldsymbol{B} $.  That is we require a field of the form
\be \label{fullFieldTwisting}
F  = \boldsymbol{E} +  \iGAj \boldsymbol{B} = \boldsymbol{E} + e_3 \boldsymbol{E} = E_0 e_1 + E_0 e_3 e_1 = E_0 e_1 + \iGAj E_0  e_2 .
\ee
Associating this twisting multivector with the electromagnetic field, we can see the conventional result of the $ \boldsymbol{E} $ and $ \boldsymbol{B} $ fields mutually orthogonal with the propagation direction. Note that if we wish the units to correspond with the conventional definition of the magnetic field $ \boldsymbol{B}' $, we need to assign $ \boldsymbol{B} \rightarrow c \boldsymbol{B}' $. 

We find, for the gradient of the field in Eq.~(\ref{fullFieldTwisting})
\bea
\partial F & = &  \left( \frac{\partial}{c \partial t} + e_3 \frac{\partial}{\partial z} \right ) \left (E_0 e_1 + E_0 e_3 e_1 \right ) \rme^{\iGAj e_3 \left ( w t - k z \right )}  \\ \nonumber
& = & \iGAj k E_0 e_1 (e_3 + 1) ( e_3  -  1 ) \rme^{\iGAj e_3 \left ( w t - k z \right )} = 0, \nonumber
\eea
where for simplicity we can ignore the $ x,y$ gradients, as they will be zero.
%Now, in order for this equation to be zero we need to add in an areal (bivector) term to the field.  That is, we can write an equation
%\be \label{NullFieldEBCondition}
%(E_0 e_1 + \iGAj c \boldsymbol{B}_0 )( e_3  -  1 ) = 0 .
%\ee
%Eq.~(\ref{NullFieldEBCondition}) can be satisfied provided the additional area term
%\be
%\iGAj c \boldsymbol{B}_0 = \iGAj e_3 \times e_1 E_0 =E_0 e_3 \wedge e_1 = \iGAj E_0  e_2.
%\ee 

%That is, for null particles at light speed, we have $ c |\boldsymbol{B}| = |\boldsymbol{E}| $, and so 
%\be
%F = \boldsymbol{E} + e_3 \boldsymbol{E} =   \left (1 + e_3 \right ) \boldsymbol{E} ,
%\ee
%where $ \boldsymbol{E} = \boldsymbol{E}_0  \rme^{\iGAj e_3 \left ( w t - k z \right )} = E_0 e_1  \rme^{e_1 e_2 \left ( w t - k z \right )} $, is the rotating electric field vector.
We have assumed a propagation direction $ e_3 $, however, for a general direction $ \hat{\boldsymbol{k}} $, we will therefore have $ \boldsymbol{k}, \boldsymbol{E}, \boldsymbol{B} $ as mutually orthogonal, as found for electromagnetic radiation.
We could also view the additional $ \boldsymbol{B} $ field as produced by a boost of the $ \boldsymbol{E} $ vector field to the speed of light, according to the conventional Lorentz transformations for the field.
%We then find the invariant
%\be
%F \cliffconj{F} = - F^2 = - \left (1 + e_3 \right ) \boldsymbol{E} \left (1 + e_3 \right ) \boldsymbol{E}  = -  \left (1 + e_3 \right )  \left (1 - e_3 \right ) \boldsymbol{E}^2 =0 ,
%\ee
%using $ e_3 \boldsymbol{E}  = - \boldsymbol{E} e_3   $, as $\boldsymbol{E} $ remains in the orthogonal $ e_1$-$e_2$ plane.
Therefore, the null particle can be described as the wave
\be \label{CircularlyPolarizedWaveMinkowski}
F = (\boldsymbol{E}_0 + e_3 \boldsymbol{E}_0) \rme^{\iGAj e_3 \left ( w t - k z \right )} =E_0 e_1 (1 - e_3  ) \rme^{\iGAj e_3 \left ( w t - k z \right )} = E_0 (e_1 + \iGAj e_2  ) \rme^{\iGAj \left ( k z - w t\right )}.
\ee
We note that we were able to replace the bivector exponential $ \rme^{\iGAj e_3 \left ( w t - k z \right )} $ with the scalar exponential $ \rme^{\iGAj \left ( k z - w t\right )} $,  due to the orthogonal nature of the propagation.  The scalar exponential will be exploited later in allowing scalar potentials to be used to represent the field.

\begin{figure}[htb]

\begin{center}
\includegraphics[width=4.4in]{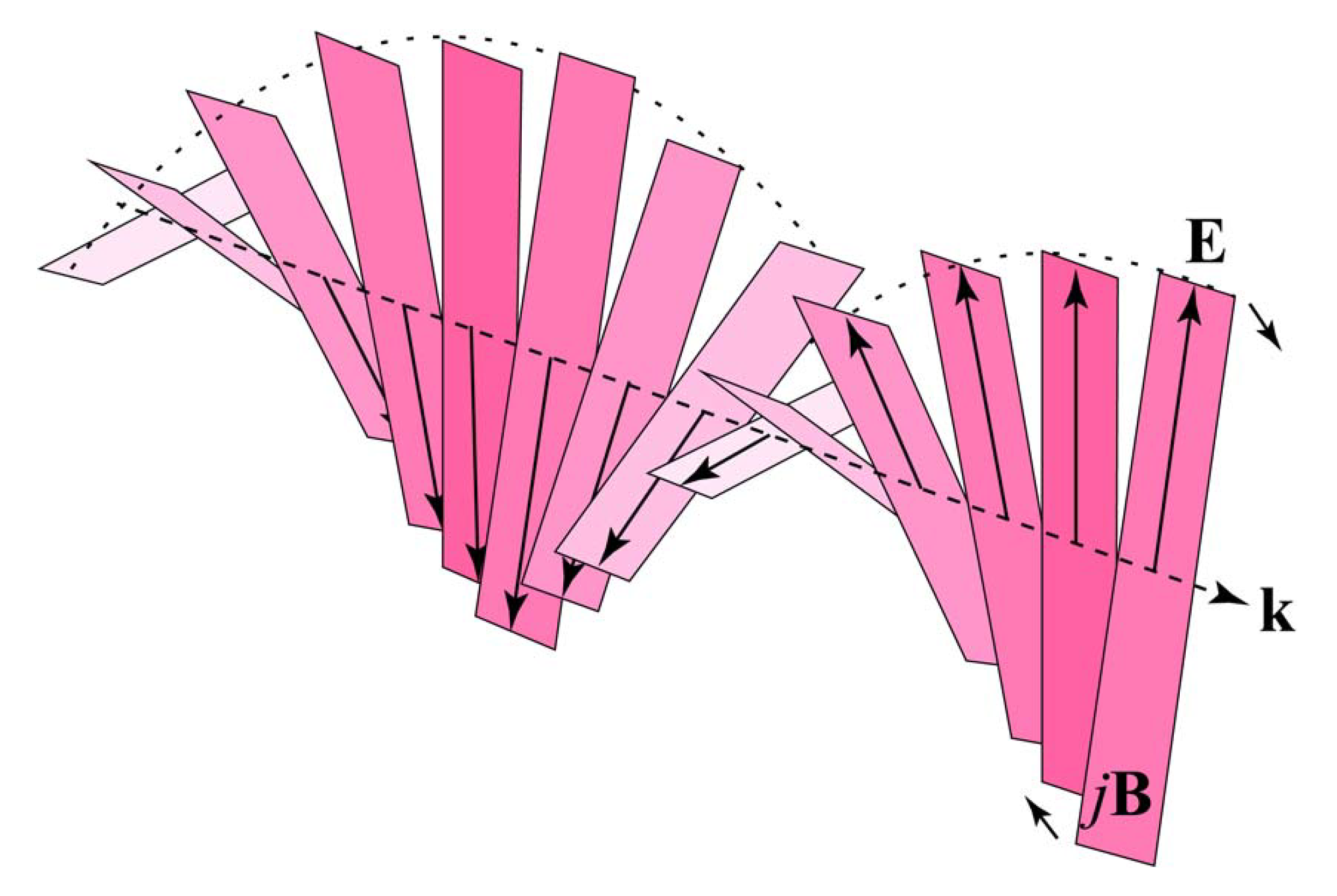}
\end{center}

\caption{A circularly polarized wave in the vacuum (left handed), in a general propagation direction $ \hat{\boldsymbol{k}} $, with the included areal term $ \iGAj c \boldsymbol{B} = \hat{\boldsymbol{k}} \wedge \boldsymbol{E} = \hat{\boldsymbol{k}}  \boldsymbol{E} $, implied either from $ F^2 = 0 $, $ \partial F = 0 $ or from a Lorentz boost of the $ \boldsymbol{E} $ field. \label{circularlyPolarizedEM}}

\end{figure}
 %Hence, by investigating the properties of null particles in $ C\ell \left ( \Re^3 \right ) $ satisfying $ \partial F = 0 $, we have produced Maxwell's equations.
%We then have
%\be
%F^2 = \boldsymbol{E}^2/c^2 -  \boldsymbol{B}^2 + 2 \iGAj  \boldsymbol{E} \cdot \boldsymbol{B}/c .
%\ee
%Hence, a null field with $ F^2 = 0 $,  requires $  |\boldsymbol{E} | = c |\boldsymbol{B} |$ and an orthogonality condition $ \boldsymbol{E} \cdot \boldsymbol{B} = 0 $, confirming our results for the circularly polarized wave. 
We can see how the geometry of null particles with $ |d X|^2 = 0 $, and fields with $ |F|^2 = 0 $, imply the conventional form of an electromagnetic wave as well as Maxwell's source free equations $ \partial F  = 0 $.
%For completeness, we note that if charge sources are present, Maxwell's equations become $ \partial F =  J $.
%, where $ J = c \rho + \boldsymbol{J} $ is a source multivector.
%The field variable $ F = \boldsymbol{E}/c + \iGAj \boldsymbol{B} $  is  isomorphic to the Riemann–Silberstein vector.

\subsection{Combining two null particles}

If define a invariant parameter $ s $, we can write
\be \label{multivectorMetricLightlike}
\frac{d X}{d s} =   \left (\frac{d t}{d s}  + \frac{d t}{d s} \frac{d \boldsymbol{x}}{d t} + \iGAj \frac{d \boldsymbol{n}}{d s} + \iGAj \frac{d b }{d s} \right ) .
\ee
Setting $ E = \frac{d t}{d s} $, $ |\boldsymbol{p}| c = E$, and with $  \frac{d \boldsymbol{x}}{d t} =  \boldsymbol{c} 
 $, and converting the angular frequency $  \frac{d \boldsymbol{n}}{d s} $ to the same energy units using the frame invariant constant $ \hbar $, we define an energy multivector
\be \label{multivectorMetricLightlikeReduced}
P = \frac{d X}{d s} =  E  + \boldsymbol{p} c + \iGAj \hbar  \boldsymbol{\omega} + \iGAj \hbar \omega .
\ee
We note that the invariance of the interval is maintained, provided the constants we use for each term is an invariant, such as $ c $ and $ \hbar $.
Therefore 
\be
|P|^2 =\frac{| d X|^2 }{ d s^2} = E^2 - \boldsymbol{p}^2 c^2 +  \boldsymbol{w}^2 - \omega^2  + 2 \iGAj \hbar \left (c \omega  -  \boldsymbol{c} \cdot  \boldsymbol{w}  \right ) = 0 ,
\ee
giving an alternate condition for null particles.
We have, $ \boldsymbol{c} \cdot  \boldsymbol{w} = c \omega  $, with $ E = p c $.

We discovered previously that each component of the multivector is separately conserved, and so if we consider two waves traveling in opposite directions, we have by summing each of the four grades
\be
 P = (E_1 + E_2) + (\boldsymbol{p}_1+\boldsymbol{p}_2) +  \hbar (\boldsymbol{\omega}_1 + \boldsymbol{\omega}_2)+ \hbar (\omega_1+\omega_2),
\ee
where $  E_1 = E_2 = E/2 $, $ \boldsymbol{p}_2 = - \boldsymbol{p}_1 $, $ \boldsymbol{\omega}_2 = - \boldsymbol{\omega}_1 $, $ \omega_2 = -\omega_1 $ and therefore $ P = E $, and therefore
\be
|P|^2 =  E^2 = m^2 .
\ee
Hence, in general a pair (or group of null particles) will have a non-zero invariant magnitude, which we will now assign as a frame invariant quantity $ m $.
Hence, in general we have for composite null particles
\be \label{magnitudeMomentumMultivector}
 |P|^2 = E^2 - \boldsymbol{p}^2 +  \boldsymbol{\omega}^2 - \xi^2  + 2 \iGAj \left ( E \xi -  \boldsymbol{p} \cdot  \boldsymbol{\omega}  \right ) = m^2,
\ee
where we have now used natural units with $ c = \hbar  = 1 $.  This thus generalizes Einstein's energy-momentum relation to include spin and helicity. We note, in general that the mass may be complex.
Hence, in general we can write for a composite object multivector
\be \label{massiveParticle}
P =  E  + \boldsymbol{p} + \iGAj \boldsymbol{\omega} + \iGAj \xi ,
\ee
where $ \omega $ can be identified with the de~Broglie frequency of a particle and $ \xi $ is the helicity, with $ |P|^2=m^2$.  We would therefore expect that the de~Broglie frequency of a pair of photons would be double the frequency of the individual photons, which is in agreement with experiments on multipohoton wave packets~\cite{fonseca1999measurement}.
This showing explicitly wave particle duality and a way of interpreting mass as composite massless photons.  It also indicates that mass is a two-dimensional complex quantity, in general.

This result in Eq.~(\ref{magnitudeMomentumMultivector}) can thus be explored as an explanation for dark matter, as we can see that mass can arise from spin and helicity, not just energy.  Also complex mass may introduce extra interaction forces that could appear as dark matter.

\subsubsection{Velocity multivector}

For a composite particle, we can now also write a generalized spacetime event $ X $, in differential form, as
\be \label{multivectorMetric}
d X = d t + d \boldsymbol{x} + \iGAj d \boldsymbol{n} + \iGAj d b = d \tau.
\ee
The magnitude can now be written
\be \label{AmplitudeSquaredProperTime}
| d X |^2   = d t^2 - d \boldsymbol{x}^2 + d \boldsymbol{n}^2 - d b^2 + 2 \iGAj \left (d b d t - d \boldsymbol{x} \cdot d \boldsymbol{n}  \right ) = d\tau^2,
\ee
 which defines the proper time. 
We note though, that in the general case, the proper time is a complex quantity and thus two-dimensional.
 
Now, dividing through by the invariant $ d \tau $, from Eq.~(\ref{multivectorMetric}), we produce the velocity multivector  
\bea \label{velocityMultivector}
V & = & \frac{d X}{d \tau} = \frac{d t}{d \tau } + \frac{d \boldsymbol{x}}{ dt }\frac{ dt }{d \tau} + \iGAj \frac{d \boldsymbol{n}}{d t} \frac{d t}{d \tau } + \iGAj \frac{d b}{d t}\frac{d t}{d \tau} \\ \nonumber
& = & \gamma \left ( 1 + \boldsymbol{v} + \iGAj  \boldsymbol{w} + \iGAj \xi \right ) , \nonumber
\eea
where $\boldsymbol{v} = \frac{d \boldsymbol{x}}{d t} $, $\boldsymbol{w} = \frac{d \boldsymbol{n}}{d t} $ and $\xi= \frac{d b}{d t}$ and $ \gamma = \frac{dt}{d \tau}$.
As we have $ | d X |^2 = d \tau^2 $ then we have $ |V|^2 = \frac{| d X |^2}{d \tau^2} = 1 $.  
Therefore
\be
|V|^2  = \gamma^2 \left ( 1 - \boldsymbol{v}^2 +  \boldsymbol{w}^2 - \xi^2  + 2 \iGAj \left (\xi -  \boldsymbol{v} \cdot  \boldsymbol{w}  \right )\right ) = 1
\ee
%= \gamma^2 \left ( 1 + \boldsymbol{v} + \iGAj  \boldsymbol{w} + \iGAj h \right )\left ( 1 - \boldsymbol{v} - \iGAj  \boldsymbol{w} + \iGAj h \right )
and hence
\be \label{fullGammExpression}
\gamma =  \frac{d t}{d \tau} = \frac{1}{\sqrt{\left ( 1 - \boldsymbol{v}^2 +  \boldsymbol{w}^2 - \xi^2  + 2 \iGAj \left (\xi -  \boldsymbol{v} \cdot  \boldsymbol{w}  \right )\right )}} ,
\ee
generalizing the time dilation factor to include spin $ \boldsymbol{w} $ and helicity $ \xi $.
%The metric in Eq.~(\ref{SpacetimeAmplitudeSquared}) thus indicates two distinct classes of particles those having a proper time with $ |d X|^2 \ne 0 $, which allows the forming of the eight-velocity, as above, commonly associated with matter particles and those with $ |d X|^2 = 0 $ described as lightlike particles.

%As already noted, $ d \tau $ is a complex-like number, in general. 

Also, as $ |V|^2 = 1 $, then it is naturally expressed in exponential form
\be
V = \rme^{\phi \frac{\boldsymbol{v} + \iGAj  \boldsymbol{w}}{\sqrt{\left (\boldsymbol{v} + \iGAj  \boldsymbol{w} \right )^2} } } = \cosh \phi +  \frac{\boldsymbol{v} + \iGAj \boldsymbol{w} }{\sqrt{(\boldsymbol{v} + \iGAj \boldsymbol{w})^2 } } \sinh \phi .
\ee
%where $ \cosh \phi = \gamma \left ( 1 + \iGAj h \right )$. 
The exponential form shows that a change in eight-velocity involves a hyperbolic rotation of the  multivector in eight dimensions.  Hence, the eight-velocity takes a dual role, of representing velocity as well as acting as an operator to change frames, according to Eq.~(\ref{HomLorentzGroup}).
%This allows the Schr\"{o}dinger wave equations and the use of differential operators to determine the particle energy and momentum.

Now, as $ |V|^2 = 1 $ is constant, then differentiating, we find $ \frac{d }{d \tau} |V|^2 = \frac{d V}{d \tau} \cliffconj{V}  + V \frac{d \cliffconj{V}}{d \tau} = 0 $, using the product rule of differentiation.  So, defining $ A = \frac{d V}{d \tau} $ for an acceleration multivector, we thus produce an orthogonality condition for the velocity and acceleration multivectors as $ A \cliffconj{V} + V \cliffconj{A} = 0 $ or $ A \cdot \cliffconj{V} = 0 $, a generalisation of the conventional four-vector result.
%As mentioned, the result that $ |V|^2 = 1 $ for all observers, implies that the magnitude of the eight-velocity is constant. Hence, the change of $V$ must be orthogonal to $V$, which is shown by $ A \cdot \cliffconj{V} = 0 $.  We have $ |A|^2 < 0 $, so that $U$ is getting a spacelike variation $ d U = A d \tau $. The spacelike acceleration, $ |A|^2 < 0 $, implies that the change to the temporal part is less than than the change to the spatial part, and so ensuring that the magnitude of $ V $ is constant. Hence, acceleration can be described as a rotation of the velocity multivector.
%In general, the spatial parts of V and A do not have to be orthogonal to each other but this is the case for circular motion with $ \boldsymbol{a} \cdot \boldsymbol{v} = 0 $, for example. 
%The velocity multivector thus also has the special property of being dimensionless.  %It also has a magnitude of zero for lightlike particles with $ |d X |^2 = 0 $, or unity for timelike particles $ |d X |^2 \ne 0 $.

\subsection{The nature of time}

When physical space is modeled with $ C\ell(\Re^3) $, producing the spacetime multivector in Eq.~(\ref{multivector3DSimple}),  time becomes identified as  the scalar grade of the multivector.
 Also, clearly time is not now a linear dimension like space, but has its own physically distinct properties.
Geometrically, the scalar property could be visualized as the radius of a spherical wavefront of a null particle, expanding at the speed of light.  As light is invariant under the Lorentz transform, this sphere will remain a sphere under boosts, and so only a single scalar quantity is necessary to define this radius.  Hence, time now essentially records the amount of expansion of the light sphere.  This is analogous to measuring time via the expansion of universe, which also creates the appearance of linear flow to time.

Now, in the general case, as shown in Eq.~(\ref{AmplitudeSquaredProperTime}), the proper time includes the pseudoscalar component of the multivector, so that time now becomes a two-dimensional quantity. 
Two-dimensional time has actually been shown to produce a viable description of nature~\cite{bars2001survey} and recently, two-dimensional time has been created within a quantum computer\cite{dumitrescu2022dynamical}. A particular version of string theory also uses a 12-dimensional spacetime having two dimensions of time, with a metric signature (10,2). It has also been shown that the wave equation with two time dimensions can be well posed and evolves deterministically~\cite{craig2009determinism}. 
This pseudoscalar aspect of proper time, is an invariant twist of space, and would relate to time measured with the helical motion of light.
Hence, these two types of time should be distinguished in physics and not conflated together.  
The second grade or bivector component of the spacetime multivector also has the same signature as time in the invariant interval, and so could also be classed as a timelike dimension.  Hence, time is more convoluted and has more subtleties than a simple dimension.  It firstly takes the property of a scalar, but also the bivector spin aspect representing three dimensions of time, as described by the hand on a clock for example, as well as the pseudoscalar element with the proper time, describing the helical nature of time, as in a photon.  Hence, we could say that time is everything that is not space, in the spacetime multivector, and overall has five-dimensions.
We could visualize two dimensional proper time as the hand of a clock having the freedom to move around the surface of a sphere and not just a circle in the plane, or perhaps more accurately as clock hands on an outwardly stretching clock face.

\subsubsection{The arrow of time}

We notice from Eq.~(\ref{AmplitudeSquared}) that an additional non-squared time factor $ dt db $ arises in the imaginary component, which breaks the symmetry of time, thus creating an arrow of time. This feature allows the avoidance of the problem of closed timelike loops with two conventional time dimensions. 
Also time as a scalar, representing a sphere of light expanding from a point, is also fundamentally a one way process.
So we can also see that if $ dt $ has a certain direction, due to the expansion of the universe, say, then in order to hold the interval invariant the helicity will have to remain the same sign. This could explain the nearly complete absence of right-handed neutrinos, handedness in molecules and the selection of matter over anti-matter.

\subsubsection{Non-local properties}

Considering Eq.~(\ref{AmplitudeSquaredProperTime}), through manipulating the spin terms, the proper time can be brought to zero, even if the speed is faster than light.  This could perhaps be a way to explain the property of entanglement.  For example, we can begin with instantaneous communication for a local measurement, with $ dx=0$ and $ dt=0$, but we can see that we can increase the separation $ dx $ provided we increase the spin $ dn $ by an equal amount, thus retaining the magnitude of the multivector.  This also implies therefore the existence of tachyons.

\subsection{The phase of waves}

We know that the expression $ P \cdot \cliffconj{X} $ will form an invariant
\bea \nonumber
 & = &  \frac{1}{2} ( E  + \boldsymbol{p} + \iGAj \boldsymbol{\omega} + \iGAj \xi) ( t -  \boldsymbol{x} - \iGAj  \boldsymbol{n} + \iGAj  b) + ( t +  \boldsymbol{x} + \iGAj  \boldsymbol{n} + \iGAj  b) ( E  - \boldsymbol{p} - \iGAj \boldsymbol{\omega} + \iGAj \xi)  \\ 
 & = & E t - \boldsymbol{p} \cdot \boldsymbol{x} + \boldsymbol{\omega} \cdot \boldsymbol{n} - b \xi + \iGAj ( E b - \boldsymbol{p} \cdot \boldsymbol{n} - \boldsymbol{x} \cdot \boldsymbol{\omega} +\xi t ),
\eea
which will be zero for null particles.
Comparing this with the expression describing  traveling waves
\be
\rme^{ i (\omega t - \boldsymbol{k} \cdot \boldsymbol{x})} ,
\ee
we can see that we have an additional two terms $ \boldsymbol{\omega} \cdot \boldsymbol{n} - b \xi$. These terms make sense in terms of a helical wave in Fig.~\ref{circularlyPolarized}, The $\boldsymbol{n} \boldsymbol{\omega}$  term shows that if we rotate the reference coordinates $\boldsymbol{n}$, then $  \boldsymbol{\omega} \cdot \boldsymbol{n} $  has an equivalent effect to moving forward in time, given by $ \omega t $.    We can make a similar argument for the helicity component $b \xi$, that is, varying the helicity $ b $, has the same effect as moving forwards in space.  In order to have a propagation that does not decay exponentially, we also need the pseudoscalar term $ \iGAj ( E b - \boldsymbol{p} \cdot \boldsymbol{n} - \boldsymbol{x} \cdot \boldsymbol{\omega} +\xi t ) $, to be zero.

The invariant interval, or distance measure, being a complex number we can write it as a magnitude and phase $  | M | = A \rme^{j \phi } $, thus leading naturally to quantum amplitudes and phases.

\section{Conclusion}

We begin with the generalised eight-dimensional Minkowski spacetime, which emerges when physical space is modelled with  $ C\ell(\Re^3) $. We find that the additional four degrees of freedom describe the properties of spin and helicity. 
Through  exploring the algebra of $ C\ell(\Re^3) $, without any reference to physical arguments, we produce a generalized Minkowski spacetime and Lorentz transformations, a generalized energy-momentum relation, the properties of electromagnetic waves and Maxwell's equation.
We also support the validity of two proposed properties in physics, of complex time and complex mass. Indeed, we unpack some of the complexity of time, with ultimately five dimensions. The three aspects of time noted are the linear expansion-type time, rotational time and finally helical time. Our generalized metric also includes a term explaining the arrow of time. This also implies a preferential handedness to the universe, as found in elementary particles and molecules, for example. 

We also show the asymmetry between electric and magnetic fields as they arise from an eight-dimensional multivector  potential, which provides a new argument against the existence of magnetic monopoles.

The generalised invariant distance, shown in Eq.~(\ref{AmplitudeSquared}), defines an inertial Lagrangian that encodes the four fundamental conservation laws, of energy, linear momentum, angular momentum and spin.  Through the minimal addition of a flow field to spacetime we reproduce gravity effects, consistent with the infalling flat space Gullstrand–Painlevé coordinates of General Relativity, as well as electromagnetic forces. 

While science has carefully explored the properties of a time dimension and spatial three-vectors within  physical space~(the scalar and vector components of the multivector), it appears to have  overlooked the basic rotational freedoms we also experience within physical space, as represented in our generalized spacetime multivector, in Eq.~(\ref{multivector3DSimple}) by including the pseudovector and pseudoscalar elements.

%\section*{Acknowledgment}

%\appendix

\section*{References}

\bibliographystyle{iopart-num}

\bibliography{main}

% ------------------------------------------------------------------------
\end{document}